\documentclass[11pt]{article}

\usepackage{amsfonts}
\usepackage{amsmath}
\usepackage{amssymb}
\usepackage{latexsym}
\usepackage{color}
\usepackage{colordvi}
\usepackage{epsfig}
\usepackage{array}
\usepackage{pifont}
\usepackage{axodraw}
\usepackage{citesort}

\newcommand{\beq}{\begin{eqnarray}}
\newcommand{\eeq}{\end{eqnarray}}

\newcommand{\be}{\begin{equation}}
\newcommand{\ee}{\end{equation}}
\newcommand{\lwrsim}{\raise0.3ex\hbox{$<$\kern-0.75em\raise-1.1ex\hbox{$\sim$}}}

\def\C2#1#2{({\cal C}_2)_{#1}^{#2}}

\def\eq#1{Eq.~(\ref{#1})}

\usepackage{axodraw,
braket
}
\newcommand{\omu}{\overline{\mu}}
\newcommand{\p}{\partial}

\newcommand{\lqcd}{\Lambda_{\mbox{\tiny{QCD}}}}
\newcommand{\msbar}{\overline{\mbox{{\sc ms}}}}
\newcommand{\VA}{\braket{A^2}}
\newcommand{\mom}{\rm{MOM}}
\def\s#1{{\scriptscriptstyle #1}}
\include{diagrams}
\title{\textbf{Nontrivial ghost-gluon vertex and the match of RGZ, DSE and lattice Yang-Mills propagators}}

\author{{\it D.~Dudal\,$^{1}$, O.~Oliveira\,$^{2}$, J.~Rodr\'{\i}guez-Quintero\,$^{3}$
}}

\date{}

\begin{document} 

\maketitle
\vspace{-10mm}
\begin{center}{\footnotesize
$^{1}$ Department of Physics and Astronomy, Ghent University, Krijgslaan 281 S9 9000 Gent, Belgium}\\
$^{2}$ {\footnotesize Departamento de F\'isica, Universidade de Coimbra, 3004-516 Coimbra, Portugal}\\
$^{3}$ {\footnotesize Dpto. F\'isica Aplicada, Fac. Ciencias Experimentales, Universidad de Huelva, 21071 Huelva, Spain}
\end{center}

\begin{abstract}
Either by solving the ghost propagator DSE or
through a one-loop computation in the RGZ (Refined Gribov-Zwanziger) formalism, we show that a non-trivial ghost-gluon vertex is anyhow required to obtain a ghost propagator prediction compatible with the available corresponding lattice data in the SU(3) case. For the necessary gluon propagator input, we present RGZ tree level fits which account well for the gluon lattice data. Interestingly, this propagator can be rewritten in terms of a running gluon mass. A comparison of both DSE and RGZ results for the ghost propagator
is furthermore provided. We also briefly discuss the connection between the RGZ and the OPE $d=2$ gluon condensate.
\end{abstract}

\vspace{-14cm}
\begin{flushright}
{\small UHU-FP/12-010}\\
\end{flushright}
\vspace{14cm}




\section{Introduction}

As the Green functions of a general quantum field theory contain in principle all information,  considerable research efforts is being devoted to construct as reliable as possible estimates of these Green functions.
Amongst the various possible ways to compute the Green functions, a powerful method is the study of the Dyson-Schwinger equations (DSEs), which are nothing else than the quantum version of the equations of motions and are thus derivable from the partition function itself \cite{Itzykson,Alkofer:2000wg,Maris2003,Binosi:2009qm,Bashir:2012fs}. The appealing it may look to actually solve the DSEs, it must be mentioned that they inevitably lead to an infinite number of coupled equations and consequently assumptions/truncations must be made to obtain any kind of results.
For  SU($N$) Yang-Mills gauge systems, usually quantized in the Landau gauge because of the special (renormalization) properties of this gauge, the DSEs for the propagators themselves are solved while the input vertices are either taken to be tree level like or modeled
using information from e.g.~lattice simulations of the non-perturbative vertices when available \cite{Alkofer2003,Aguilar:2004sw,Aguilar:2008xm,Boucaud:2008ky,Fischer:2008uz,Binosi:2009qm,Aguilar:2010cn,RodriguezQuintero:2010wy,Pennington:2011xs,Bashir:2011dp}.

The Landau gauge is also perfectly suited for a numerical lattice approach, allowing for explicit comparisons between continuum results and their lattice counterparts (for a recent review see \cite{Boucaud:2011ug}), which are a priori non-perturbative and exact, apart from the usual difficulties of correctly handling discretization artifacts~\cite{Becirevic1999,Becirevic2000,OliveiraSilva2005,Soto2007,Oliveira:2012}. This means certain assumptions can be tested, for example if it is allowed to invoke for certain vertices the tree level value.
The here presented work fits herein. Specifically, we will discuss the solution of the ghost propagator DSE (GPDSE), using a fine modeling of the input gluon propagator from precise large volume lattice data.
The variable ingredient of the DSE will be the ghost-gluon vertex.

In a previous paper \cite{Boucaud:2011eh}, some of us already discussed how to model the ghost-gluon vertex inspired by an Operator Product Expansion (OPE) analysis with a $d=2$ condensate, allowing for a reasonable description of the available vertex' lattice data. With the exception of some (semi-classical) analysis' of the vertex made in \cite{Schleifenbaum:2004id,Pennington:2011xs} and the very recent $d=2$ work of \cite{Huber:2012zj}, other DSE works usually deploy a tree level ghost-gluon approximation. As advocated in several independent series of DSE papers, \emph{qualitative} agreement between the gluon/ghost lattice data is found with what is now known as the massive gluon (decoupling) DSE solution ({\it i.e.}, see \cite{Aguilar:2008xm,Boucaud:2008ky,Fischer:2008uz,Binosi:2009qm}). This {\it qualitative agreement} should be understood as that the coupling at the renormalization point is rather treated as a fitting parameter in order to account for the lattice data. Otherwise, mismatches between DSE and lattice results appear, as it is particularly manifest for the ghost dressing function~\cite{Boucaud:2008ji}. It should be remembered that, after the DSEs have been appropriately renormalized, the renormalized coupling could also be taken from the lattice as an input, in which case clearly is not an additional fitting parameter. Although such a qualitative agreement may be the first concern, once this is settled one should also aim at agreement at the \emph{quantitative} level, otherwise we cannot claim that the DSEs capture all relevant dynamics.

In this paper, we set an important step in this program. We shall show that a tree level (bare) ghost-gluon vertex never leads to an acceptable ghost DSE estimate of the lattice ghost propagator. Such could already have been guessed from the results of studies in this approximation, in which case there was always a mismatch with the lattice ghost (see e.g.~\cite{Aguilar:2008xm,Boucaud:2008ji,RodriguezQuintero:2011vw}). Incorporating a nontrivial vertex model, compatible
with the lattice data, based on the preparatory work \cite{Boucaud:2011eh}
does lead to a fine match. Simplified versions of this vertex model also keep leading to discrepancies in the lower momentum regime, as well do the analytical one-loop approximations to the ghost propagator using the Refined Gribov-Zwanziger formalism,  again using the latter simplified models. This clearly illustrates the important role played by the ghost-gluon vertex to adequately describe the Landau gauge ghost propagator within the DSE approach.

In what follows, we shall first describe our setup and used input, before coming to results and our conclusion.

\section{A precise modeling of the input gluon propagator}
\label{sec:gluon}
The main motivation of this paper is the analysis of the ghost dressing results obtained from the
ghost DSE, particularly in connection with the input for the ghost-gluon vertex. The first step should
be to provide ourselves with a proper and reliable model for the gluon propagator, which accounts for the corresponding ``ab initio'' QCD lattice data. As will be
discussed below, this will permit us to solve the ghost DSE (in isolation from the gluon DSE) with
the ghost-gluon vertex, a main target of this study, as the only unknown ingredient, and finally
compare with available ghost lattice data.

\subsection{The RGZ gluon propagator and a running gluon mass}
Since the key work of Gribov \cite{Gribov:1978kw}, it has become well-appreciated that the Landau gauge does not allow to pick out a single gauge field per orbit: multiple gauge equivalent fields all fulfilling $\partial_\mu A_\mu=0$ can exist. As a consequence, the usual Faddeev-Popov procedure becomes in principle inadequate at the non-perturbative level. The Gribov copy problem is indeed not expected to play a role in the ultraviolet where perturbation theories applies due to asymptotic freedom. However, the infrared dynamics is severely changed.

In the continuum, the up-to-date best worked out (partial) cure to the Gribov problem ---at least to our knowledge--- is based on the original Gribov-Zwanziger approach: when the functional $\mathcal{R}[A]=\int d^4x A^2$ is minimized per gauge orbit, a necessary condition for an extremum is the transversality condition $\p_\mu A_\mu=0$, i.e.~the Landau gauge. A \emph{local} minimum is approached when the Faddeev-Popov operator, $-\p_\mu D_\mu$, is positive\,\footnote{This statement makes sense in the Landau gauge in which case the Faddeev-Popov operator is Hermitian.}. The latter requirement is closely related to the occurrence of Gribov copies linked with zero modes of the Faddeev-Popov operator, as it is immediate to recognize that the existence of a mode $\omega$ with $-\p_\mu D_\mu \omega=0$ leads to the existence of an (infinitemisal) gauge copy due to $\p_\mu\left[A_\mu+\epsilon D_\mu\omega\right]=\p_\mu A_\mu$. Therefore, as a first (and to date the only practically worked out) way to at least kill off this subset of infinitesimally related gauge copies in the continuum world, Gribov and Zwanziger implemented the restriction of the gauge path integral to the so-called Gribov region, where the Faddeev-Popov operator is positive\,\footnote{In the lattice approach to Landau gauge dynamics, the Gribov problem is also there, see e.g.\,\cite{Silva:2004} and \cite{Hughes:2012hg} for a recent toy model analysis, and it is usually treated in a quite similar fashion: one restricts the gauge fields admissible to the Monte Carlo sampling to a subset belonging to the first Gribov horizon, i.e.\,one numerically searches for the ``optimal'' minima of the Landau gauge functional.}. Let us refer to the original works \cite{Gribov:1978kw,Zwanziger:1989mf} and the recent overview \cite{Vandersickel:2012tz} for the underlying details. For the purpose of this paper, it is sufficient to know that this restriction can be brought into the working form of a new partition function, entailing the presence of a non-perturbative mass (Gribov) parameter $\gamma^2\propto \lqcd^2$. Formally, for $\gamma^2=0$, the conventional Faddeev-Popov action is recovered, so it becomes evident that the Gribov copies' effect, as carried by $\gamma^2$, is a pure non-perturbative effect, and the presence of the ``soft'' mass parameter $\gamma^2$ is only influencing the infrared structure of the theory. For example, the predicted tree level gluon propagator was
\begin{equation}\label{eq:GZ}
D(p^2)\ = \  \frac{p^2}{p^4+2g^2N\gamma^4} \,.
\end{equation}
Due to the apparent mismatch of this propagator with state-of-the-art lattice data of e.g.~\cite{Cucchieri:2007md,Bogolubsky:2007ud,Sternbeck:2007ug,Cucchieri:2007rg,Maas:2008ri,Bogolubsky:2009dc,Oliveira:2011zn}, it became clear this was not the end of the story. In a series of papers \cite{Dudal:2005na,Dudal:2007cw,Dudal:2008sp,Dudal:2011gd}, it has been discussed that the Gribov-Zwanziger action dynamically improves itself by the condensation of dimension 2 condensates, including the elsewhere much studied gluon condensate $\VA$ ({\it e.g.}, see \cite{Boucaud:2000nd,Gubarev:2000nz,Kondo:2001nq,Verschelde:2001ia,Dudal:2002pq,RuizArriola:2004en,Vercauteren:2011ze}).
The tree level gluon propagator of this Refined Gribov-Zwanziger (RGZ) scheme is provided by
\begin{equation}\label{eq:RGZtreeD}
D(p^2)\ = \  \frac{p^2+M^2}{p^4+(M^2+m^2)p^2+\lambda^4} \ ,
\end{equation}
where the parameters $M^2$ and $m^2$ are related to the local condensates of the gluon and
auxiliary fields intended to preserve locality of the RGZ action \cite{Dudal:2007cw,Dudal:2008sp,Dudal:2011gd}. In particular,
\beq\label{eq:m2A2}
m^2 \ = \ - \frac{13 N_C}{9(N_C^2-1)} g^2 \langle A^2 \rangle_\s{\rm RGZ} \ ,
\eeq
where $g^2 \langle A^2 \rangle_\s{\rm RGZ}$ is the dimension-two condensate of the local operator $A^2$, also appearing
in the OPE analysis of lattice Green functions \cite{Boucaud:2000nd,Boucaud:2001st,Boucaud:2005xn,Boucaud:2008gn}.
The renormalization prescription for this local operator followed in OPE or RGZ approaches may differ and,
as will be discussed below, this would imply that the non-perturbative condensates are differently defined.
The $\lambda$ parameter in \eq{eq:RGZtreeD} can be also related to the Gribov
parameter $\gamma^2$,
\beq
\lambda^4 \ = \ m^2 M^2 + 2g^2 N_c \gamma^4 \ .
\eeq
\eq{eq:RGZtreeD} has been successfully applied to describe SU(3) and SU(2) lattice gluon propagators \cite{Dudal:2010tf,Cucchieri:2011ig}. We shall therefore also use it here to model in an adequate fashion the
MOM-renormalized gluon propagator. It is interesting to notice that, as also done in \cite{Weber:2012vf},
the propagator \eqref{eq:RGZtreeD} can be reinterpreted as
\beq\label{eq:gluonR}
D^{\mom}(k^2) \ = \ \frac{z(\mu^2)}{k^2+\overline{m}^2(k^2)} \ ,
\eeq
in terms of what we can call an effective ``running RGZ gluon mass''\footnote{For a discussion on the running gluon
mass and its modelling see
e.g.~\cite{Cornwall:1981zr,AguilarPapavassiliou2010mass,OliveiraBicudo2011,Pennington:2011xs}.},
\beq\label{eq:m2}
\overline{m}^2(k^2) \ = \ \frac{m_0^2}{1+\displaystyle \frac{k^2}{M^2}} -
\frac{13 N_C}{9(N_C^2-1)} \frac{k^2}{k^2+M^2} g^2 \langle A^2 \rangle_\s{\rm RGZ}
\eeq
with
\beq\label{eq:m0}
m_0^2 = \frac{2g^2N_C\gamma^4}{M^2} - \frac{13 N_C}{9(N_C^2-1)} g^2 \langle A^2 \rangle_\s{\rm RGZ} \ ,
\eeq
and where $\mu^2$ is the renormalization momentum at which the MOM prescription determines that
\beq
z(\mu^2) \ = \ 1 + \frac{\overline{m}^2(\mu^2)}{\mu^2} \ .
\eeq
It should be noticed that the mass function in \eq{eq:m2} is gifted (as can be also seen in
Fig.~\ref{fig:bestpar}) with the main feature of reaching a positive saturation point (frozen mass), at asymptotically low momenta,
that shifts toward a negative correction when the momentum increases,
\beq\label{eq:m2asym}
\overline{m}^2(k^2) \ = \ \left\{
\begin{array}{lr}
m_0^2 & \text{for }k^2 \ll M^2 \\
\displaystyle
- \frac{13 N_C}{9(N_C^2-1)} g^2 \langle A^2 \rangle_\s{\rm RGZ} & \text{for }
k^2 \gg M^2
\end{array}
\right. \ .
\eeq
This feature has also been ---at least qualitatively--- observed in the results from
lattice simulations. In particular, the OPE power corrections to the
perturbative prediction for the gluon propagator, that have been shown as
unequivocally needed to describe the lattice data in refs.~\cite{Boucaud:2000nd,Boucaud:2001st},
appears to behave as \eq{eq:gluonR} with \eq{eq:m2asym} and
dominated at intermediate momenta by the dimension-two gluon condensate of the same local operator.
This is a qualitatively striking result, although, as will be seen below,
the condensates in RGZ and OPE approaches might have been differently defined.

Then, \eq{eq:gluonR} is a good candidate to model the gluon propagator and will be confronted
to SU(3) lattice data in the next subsection. Before this, a few words about how the condensates
are defined and can be related in both approaches might be here in order.
On should remember that, in the OPE analysis, one would be left
with
\beq\label{eq:DOPE}
k^2 D^{\rm MOM}(k^2) \ = \ c_0\left(\frac{k^2}{\mu^2},g\right) \
\left( 1 +c_2\left(\frac{k^2}{\mu^2},g\right) \
\frac{\langle A^2 \rangle_\s{\rm OPE}}{4(N_C^2-1)k^2} + \ {\cal O}\left(1/k^4\right) \ \right)
\eeq
where all the quantities are assumed to be renormalized at the subtraction point $\mu^2$ and,
at tree-level, one obtains \cite{Boucaud:2000nd}
\beq
c_0\left(\frac{k^2}{\mu^2},g \right) \ = \ 1 \ , \ \ \ \ \
c_2\left(\frac{k^2}{\mu^2},g \right) \ = \ N_C g^2 \ .
\eeq
The local operator $A^2$ contains a quadratic UV divergence that should be properly
subtracted (see, for instance, the discussion of ref.~\cite{Boucaud:2002jt}),
for instance, by introducing a normal product referred to the QCD
perturbative vacuum. In particular, the non-perturbative condensate in \eq{eq:DOPE}
should be specifically understood such that, given a l.h.s.~that is non-perturbatively computed
({\it e.g.}, on a lattice) and the Wilson coefficients $c_0$ and $c_2$ computed in perturbation
at a certain order and for a particular renormalization scheme, as the product of
$c_2$ and the condensate should be free of divergencies\,\footnote{In the Shifman-Vainshtein-Zakharov
sum-rules approach~\cite{Shifman:1978bx,Shifman:1978by}, the UV divergences for the local operator and the IR ones coming from the Wilson coefficient cancel each other.}, $\langle A^2 \rangle$ is unambiguously defined.

In the underlying effective potential approach to $\VA_\s{\rm RGZ}$, the divergences of $A^2$ are consistently been taken care off using an order by order multiplicative renormalization \cite{Verschelde:2001ia,Dudal:2002pq,Dudal:2011gd}. As discussed in \cite{Verschelde:2001ia}, this approach allows to investigate non-perturbative UV contributions to $\VA$. If $\VA$ is not a purely infrared-physics-dominated condensate, part of it escapes the usual OPE dictionary. This observation reinforces the fact that $\VA_\s{\rm RGZ}$ and $\VA_\s{\rm OPE}$ are not necessarily the same objects\,\footnote{As there was some recent controversial work about the notion of condensates (see~\cite{Brodsky:2012ku,Lee:2012ga} and references therein), we can only mention  the condensates in the current work have to be understood as mass-scale parameters respectively related to the gluon propagator OPE expansion or the RGZ effective potential. Trying to properly settle the connection of both condensates
(and check this from lattice results) remains as an interesting open question.}.

Since the exact computation of the effective potential is out of the question, as is thus the exact determination of the condensate(s), one can try to use the lattice data as a guide towards the value of e.g.~$\VA_\s{\rm RGZ}$. We can also expand the RGZ gluon propagator from \eq{eq:gluonR} at large momenta
in powers of $1/k^2$ and,  after comparing with \eq{eq:DOPE} at tree-level, we will be left with
\beq\label{eq:relA2}
g^2 \langle A^2 \rangle_\s{\rm OPE} \simeq \frac{52} 9 g^2 \langle A^2 \rangle_\s{\rm RGZ} \ .
\eeq
This last result, grounded in the tree-level RGZ and OPE
expansions, has to be taken of course as a rough
approximation, but it already gives a clear indication that both
condensates are differently defined, and do play a different role. In the OPE approach $\langle A^2 \rangle_\s{\rm OPE}$ serves to incorporate non-perturbative infrared effects in the \emph{high(er) momentum} regime, while $\langle A^2 \rangle_\s{\rm RGZ}$ and the other RGZ condensates are relevant for the \emph{low(er) momentum} regime. As we will mainly focus on the tree level RGZ propagator here, we are clearly omitting any log contributions which are known to be important once the momentum gets larger\,\footnote{At low momentum, the mass scales will shield the Landau pole, allowing for a ``freezing'' of the logs.}. So, we should not expect a tree level RGZ propagator to fit the whole momentum region. We shall come back to this subtle issue shortly.

\subsection{The SU(3) lattice gluon propagator}

In the present work, we will borrow the lattice data simulated at very large lattices
(lattice lengths ranging from around 11 to 16 fm) by the Berlin-Moscow-Adelaide group~\cite{Bogolubsky:2009dc}. Of all available volumes, we will consider only those
where data is available both for the gluon and ghost propagators.

The lattice QCD simulations were performed on a finite 4D torus using the Wilson action.
For pure Yang-Mills simulations one can take the lightest glueball mass $\approx 1.7$ GeV
as a typical hadronic scale, the corresponding length scale being $L \approx 0.1$ fm.
In order to investigate the infrared properties of the theory one should use sufficiently large volumes
to avoid the problem of finite size effects and a resolution which accommodates the typical length
scales for pure Yang-Mills theory. In practice, we have to compromise between using a large physical
volume and a minimal length defined for the lattice. For the data sets considered here, the physical
volume is sufficiently large to expect essentially no  finite volume effects. However, the lattice
spacing $a \approx 0.17$ fm is about twice the typical hadronic length scale and one can expect
finite size corrections due to the relative large $a$. For the gluon propagator
the analysis of the combined effects of finite
volume and finite lattice spacing is described in \cite{Oliveira:2012}.
The ensembles considered in this work are described in Tab.~\ref{LatticeSetup}.

\begin{table}[t]
   \begin{center}
   \begin{tabular}{l@{\hspace{0.5cm}}l@{\hspace{0.3cm}}l@{\hspace{0.3cm}}r@{\hspace{0.3cm}}r@{\hspace{0.3cm}}r@{\hspace{0.3cm}}r}
      \hline
      $\beta$  &  $a$  & $~ ~ 1/a$     & $N$ & $La$ &  Conf \\
                    &  (fm)    & (GeV)  &         & (fm)    &   \\
      \hline
      5.7    & 0.1702 & 1.1595   & 80  & 13.62 & 25  \\
               &             &               & 64  & 10.89 & 14  \\
      \hline
   \end{tabular}
   \caption{\small Lattice setup. The number of configurations refers to those used to
                             compute the gluon propagator. For the ghost propagator on the $80^4$
                             lattice only 11 configurations were considered.
                            The physical scale was set using $r_0 = 0.5$ fm.
                            We refer to  \cite{Bogolubsky:2009dc} for more details.}
    \label{LatticeSetup}
   \end{center}
\end{table}

If we apply a MOM renormalization prescription at the subtraction point $\mu^2$, we get
\beq
 D^{\rm MOM}(p^2)   \ = \ \frac{D^{\rm Latt}(p^2)}
 {Z_3^\s{\rm MOM}(\mu^2)} \
\ = \ \frac{1}{\mu^2}
  \label{Eq:glueMOM}
\eeq
where $D^{\rm Latt}(p^2)$ is the bare lattice propagator and the renormalization constant
$Z_3^\s{\rm MOM}$ is non-perturbatively defined as
\beq
\left. D^{\rm MOM}(p^2) \right|_{p^2 = \mu^2}    \ = \ \frac{D^{\rm Latt}(\mu^2)}
 {Z_3^\s{\rm MOM}(\mu^2)}
\ = \ \frac{1}{\mu^2}  \ .
\eeq
This last renormalization prescription is
required (\ref{Eq:glueMOM}) to be fulfilled at the subtraction point
$\mu = 3.6$ GeV.

\begin{figure}[t]
\begin{center}
\begin{tabular}{cc}
\includegraphics[width=7.75cm]{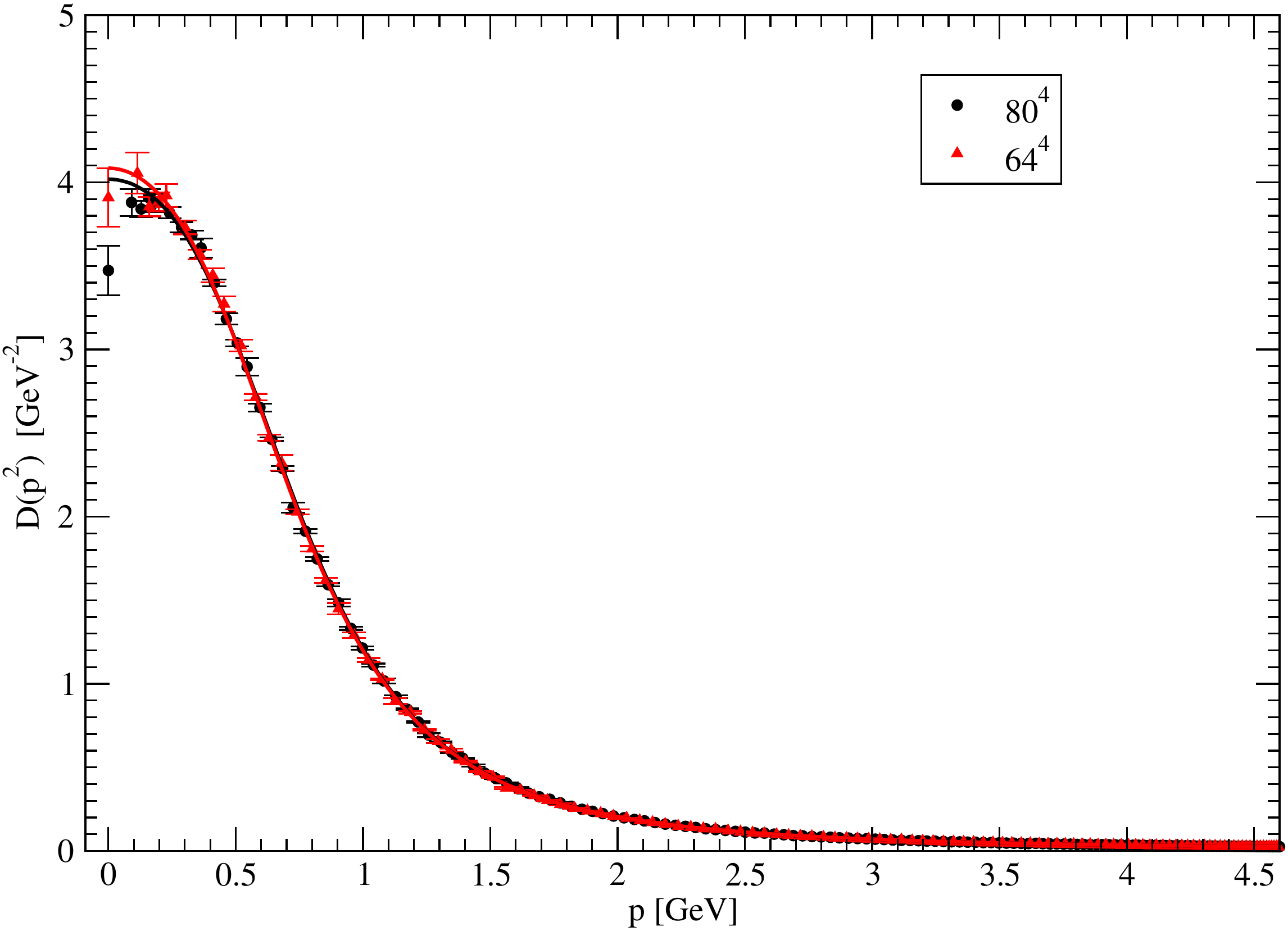}  &
\includegraphics[width=7.75cm]{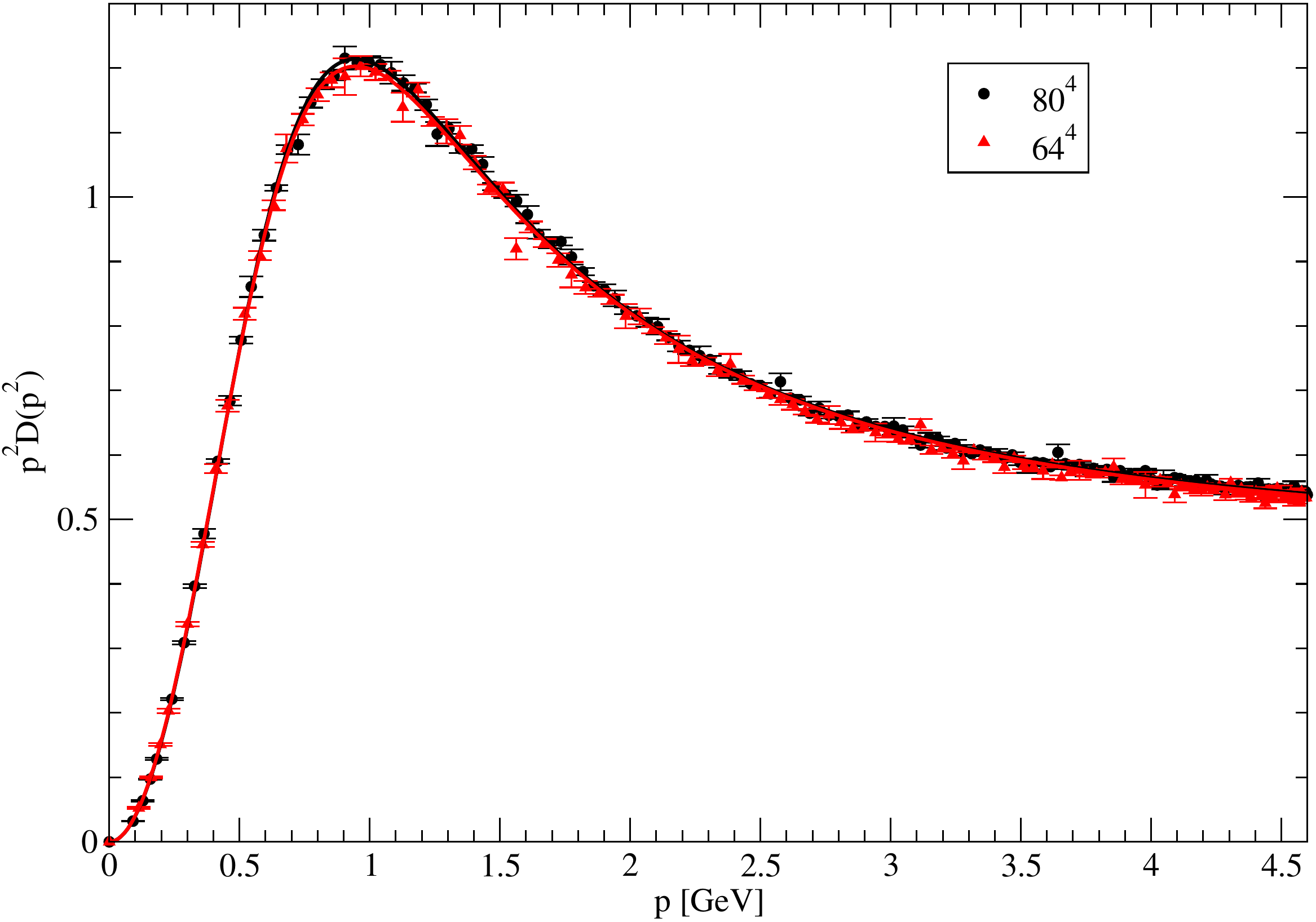}
\end{tabular}
\end{center}
\caption{\small Renormalized gluon propagator $D(p^2)$ (left) and gluon dressing function $p^2D(p^2)$ (right), the lattice data and the corresponding fits (solid lines) to \eq{newfit} using
the results reported in Tab.~\ref{tab:pmax4}. }
\label{fig:bestfit}
\end{figure}

\begin{figure}[t]
\begin{center}
\begin{tabular}{cc}
\includegraphics[width=7.75cm]{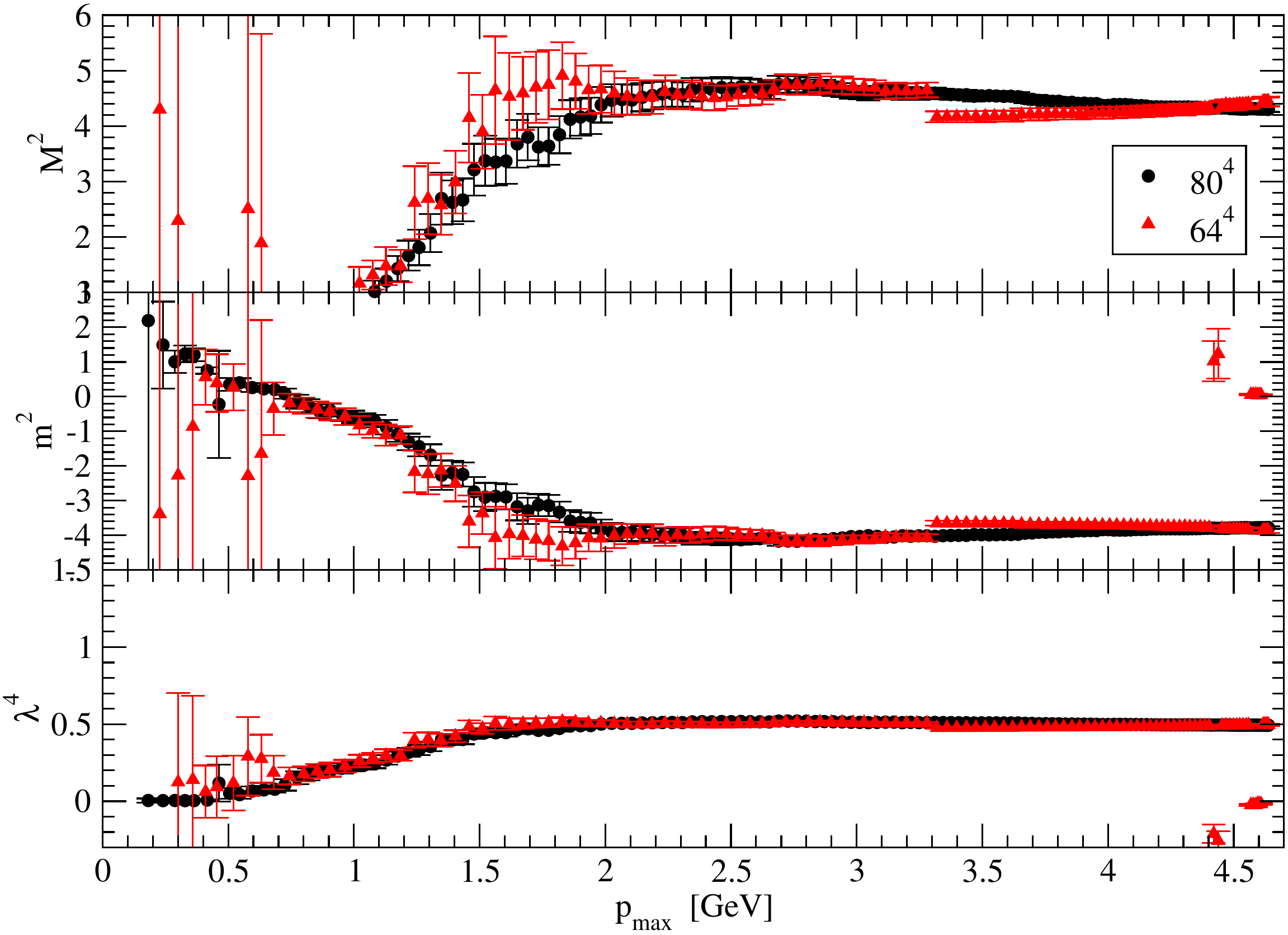} &
\includegraphics[width=7.75cm]{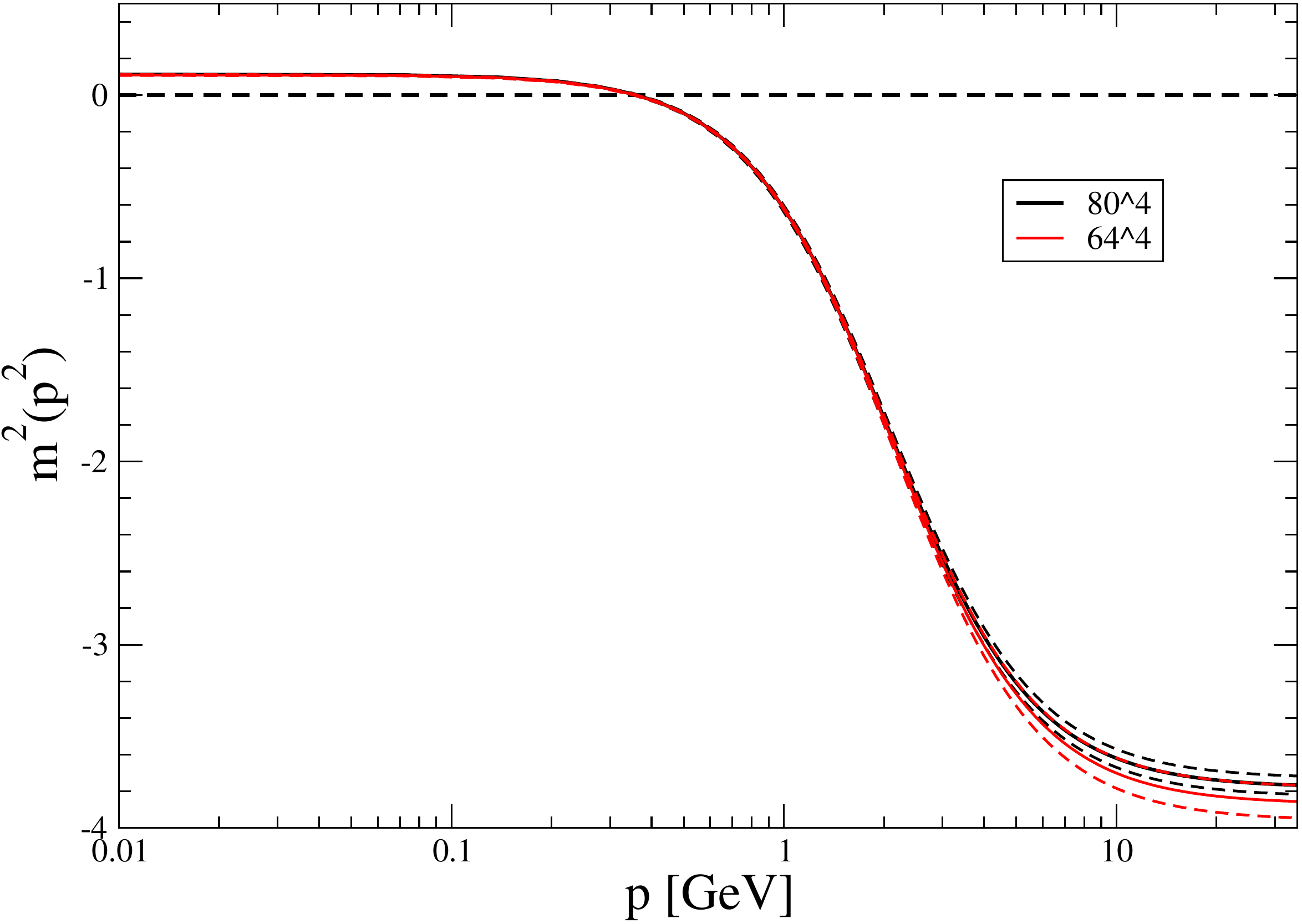}
\end{tabular}
\end{center}
\caption{\small (Left) The value of fitting parameters obtained for $p<p_{\s{\rm max}}$,
plotted in terms of $p_\s{\rm max}$. (Right) the RGZ mass function given by \eq{eq:m0} for
the parameters of Tab.~\ref{tab:pmax4}. The dotted lines account for the statistical errors to define the 1-sigma
allowed region. The ``tachyonic'' contribution, resulting from the gluon condensate, is clearly dominating the
mass function above roughly 0.5 GeV.}
\label{fig:bestpar}
\end{figure}

As the main purpose of this letter is to investigate the ghost propagator Dyson-Schwinger equation,
we need an as good as possible fit to the lattice gluon data over a so wide as possible momentum
regime. The free RGZ-model inspired formula in \eq{eq:gluonR}  can do this job. This is achieved by
introducing a global rescaling factor $Z$ as an additional fitting parameter for the tree-level RGZ
propagator given by \eq{eq:RGZtreeD},
\begin{equation}\label{newfit}
    D(p^2) \ = \ Z \ \frac{p^2+M^2}{p^4+(M^2+m^2)p^2+\lambda^4} \ .
\end{equation}
The renormalized gluon propagator is then fitted to this last equation
up to momenta $p\leq 4.638~\text{GeV}$ and the results can be found in Tab.~\ref{tab:pmax4} and
seen in Fig.~\ref{fig:bestfit}, where the lattice data and fitted curves appear displayed.
We call the reader's attention to the good agreement for the fitted parameters from the
two lattice volumes. This should be taken as a strong indication that finite-size lattice
artifacts appear to play a not very important role in our present analysis.  In addition, in the left plot of Fig.~\ref{fig:bestpar} we display the variation of the fitting parameters $m^2$, $M^2$ and $\lambda^4$ as a function of the fitting range, i.e.~obtained from fitting the data up to a maximum value of the momentum, $p_\s{\rm max}$. We clearly observe a stability in the parameters' value from $p\simeq 2\,\text{GeV}$ on.

\begin{table}[htb]
   \begin{center}
   \begin{tabular}{l@{\hspace{0.5cm}}l@{\hspace{0.3cm}}l@{\hspace{0.3cm}}l@{\hspace{0.3cm}}l
  | @{\hspace{0.3cm}}l  | @{\hspace{0.3cm}}l}
   \hline
   N  & $Z$ & $~M^2$       &  $m^2+M^2$    & $\lambda^4 $  & $m_0$  & $\chi^2/d.o.f.$ \\
      &   & (GeV$^2$)     & (GeV$^2$)  & (GeV$^4$)      &   (MeV)   &       \\	
   \hline
   80 &   0.7838(17) & 4.303(50)  & 0.526(14) & 0.4929(39) & 338(3) & 1.31 \\
   64 &   0.7800(29) & 4.442(86)  & 0.576(24) & 0.4964(64) & 334(4) & 1.72 \\
   \hline
   \end{tabular}
   \end{center}
      \caption{\small Fitted parameters for \eq{newfit} to the lattice data with the setup as in Tab.~\ref{LatticeSetup} and the
      frozen RGZ gluon mass, $m_0$, defined by \eq{eq:m0}.}
    \label{tab:pmax4}
\end{table}

\subsection{A few words about the $\langle A^2 \rangle$ condensate}
Let us stress again here that the fit \eqref{newfit} is not exactly a tree level RGZ prediction, since $Z\neq 1$. Indeed, at tree level perturbation theory, we should have in principle $Z^{\rm MOM}=1$, viz.~the first order term of $Z^{\rm MOM}=1+\ldots g^2+ \ldots g^4+\ldots$.  We could of course drop the $Z$ by reestablishing the appropriate MOM subtraction scale via solving
\begin{equation}\label{MOM}
    \frac{\mu^2+M^2}{\mu^4+(M^2+m^2)\mu^2+\lambda^4}=\frac{1}{\mu^2}
\end{equation}
with the fitted values of Tab.~\ref{tab:pmax4}. Since we are fitting quite deep into the UV here, the extracted value for $\VA_\s{\rm RGZ}$ (or any other condensate) might emulate some of the already missing log information and as a consequence, become ``unnaturally'' large. Using the numbers of Tab.~\ref{tab:pmax4} and the connection \eqref{eq:m2A2} we find
\begin{equation}\label{est}
    \braket{g^2A^2}_\s{\rm RGZ}\simeq 7~\text{GeV}^2
\end{equation}
This estimate can be compared with the tree level OPE value of \cite{Boucaud:2008gn} that, obtained at a typical deep UV scale of
10 GeV, gives $\simeq5(1)$ GeV$^2$. The two aforementioned condensate estimates imply a serious violation of the roughly derived rule \eqref{eq:relA2}, suggestive of important missing information. In particular, as we applied a tree-level RGZ formula to
describe the lattice data and extract the condensate value, its natural renormalization scale should be the one obtained by solving \eq{MOM}, $\simeq0.4$ GeV,  that lies deeply in the IR and makes questionable to use the renormalization group equation (RGE)
to run the subtraction point from the OPE scale down to this momentum. Let us furthermore
remark here that including RGE logarithms into Wilson coefficients during the OPE analysis is also of importance for a reliable determination of $\VA_\s{\rm OPE}$ (see, for instance, refs.~\cite{Boucaud:2001st,Blossier:2010ky}).

For the rest of this paper, we shall not need to worry anymore about the subtleties of interpreting the fitting parameters in terms of condensates. We can simply take the high quality fit \eqref{newfit} for $p\leq 4.638~\text{GeV}$, see Tab. \ref{tab:pmax4}, and feed this into the GPDSE.  Furthermore, as no important finite-size effect have been observed (the fitted parameters are compatible once their statistical errors are considered), we shall apply in the following the fit corresponding to data simulated at the largest lattice volume. 

\section{The ghost propagator DSE, its 1st iteration (one-loop perturbation theory),
full numerical solution and the capital issue of the ghost-gluon vertex}

The goal motivating this section is the computation of the ghost dressing function
by solving the ghost gap equation,
\begin{equation}\label{ghgl0}
F(k^2) \, = \, \frac{1}{1- \sigma(k^2)} \; ,
\end{equation}
where $\sigma(k^2)$ is the ghost self-energy that reads in $d$ dimensions
\begin{equation}\label{ghgl1}
\sigma(k^2) \, = \, g^2 N_C \frac{k_{\mu} k_{\nu}}{k^2} \int \frac{d^d q}{(2\pi)^d} H_1(k-q,k)\frac{F((k-q)^2)}{(k-q)^2}
D(q^2) \left(\delta_{\mu\nu}-\frac{q_\mu q_\nu}{q^2}\right) \ ,
\end{equation}
the form factor $H_1(k-q,k)$ is non-longitudinal one in the ghost-gluon vertex,
\beq
\widetilde{\Gamma}_\nu^{abc}(q-k,k;-q) \ &=& \ i g_0 f^{abc} q_{\nu'}
\widetilde{\Gamma}_{\nu'\nu}(q-k,k;-q) \nonumber \\
&=&
i g_0 f^{abc} \left[ \ (k-q)_\nu H_1(k-q,k) - q_\nu H_2(k-q,k) \ \right] \ ,
\label{DefH12}
\eeq
$k-q$ and $k$ are respectively the outgoing and incoming ghost momenta and $g_0$
is the bare coupling constant. Now, the RGZ gluon propagator discussed in the previous section
will be taken for $D(q^2)$ and applied to solve \eq{ghgl0} either analytically, by taking
$F(q^2)=1$ under the integral (as a first perturbative iteration), or by iterating numerically
until a satisfactory convergence of results is reached.

\subsection{Analytical approach to the ghost dressing function}\label{subsec:anal}

We will take here $F(q^2)=1$ in the ghost self-energy of \eq{ghgl1} and rewrite the RGZ
gluon propagator obtained in the previous sections as
\begin{equation}\label{ghgl2} D(p^2)\ = \ z_0 \ \frac{p^2+M^2}{p^4+(M^2+m^2)p^2+\lambda^4} \ = \ z_0 \ \left(\frac{\alpha_+}{p^2+\omega_{+}^2}
+\frac{\alpha_-}{p^2+\omega_{-}^2}\right) \ ,
\end{equation}
where the poles are assumed to be $cc$, with positive real part.  It has already been illustrated in \cite{Cucchieri:2011ig} (SU(2) case) and  \cite{Dudal:2010tf,Oliveira:2012} (SU(3) case) that these forms allow for a consistent fit to large volume lattice data. Propagators with pairs of $cc$ poles were already considered earlier in a DSE approach, see \cite{Stingl:1994nk}.

Now, on the ground of the OPE analysis performed in ref.~\cite{Boucaud:2011eh}, we model the non-longitudinal
form factor, $H_1(q,k)$ with the simplified non-perturbative formula:
\begin{equation}\label{ghgl3}
    H_1(q,k) \ \simeq H_1(q,0) \ \equiv W(q^2) \ = \ 1+\frac{a^2q^2}{b^4+q^4} \ ,
\end{equation}
based, as will be discussed below, on the approximation of a very soft dependence on the incoming ghost
momentum, where $b$ is a mass dimension scale discussed in ref.~\cite{Boucaud:2011eh} and
\beq\label{eq:a2A2}
 a^2 \ = \ \frac{N_C}{4 (N_C^2-1)} \ g^2 \langle A^2 \rangle_\s{\rm OPE} \ .
\eeq
We expect that, even in the approximation of a soft dependence on the incoming momentum, the simple model vertex \eqref{ghgl3}
can capture the non-perturbative essentials of the ghost-gluon vertex, while still allowing for analytical computations.
In the numerical analysis performed in the next subsection, we will discuss a more general
formula that has been derived in ref.~\cite{Boucaud:2011eh} to describe more properly the ---albeit not so precise---
current lattice data~\cite{Maas:2006qw}. Here we follow an approach, developed in \cite{Cucchieriprogress} to study one-loop RGZ predictions for the ghost propagator, and as such we only summarize the underlying computation. The result with the tree-level vertex ($W(q^2)=1$) has been previously obtained in the one-loop computation in
ref.~\cite{Cucchieri:2012cb}. The one-loop contribution to the ghost self-energy coming from the model vertex \eqref{ghgl3} yields
\begin{equation}\label{ghgl4}
\frac{k_{\mu} k_{\nu}}{k^2} \int \frac{d^d q}{(2\pi)^d} (W((k-q)^2)-1) \frac{1}{(k-q)^2}  D(q^2)
\left(\delta_{\mu\nu}-\frac{q_\mu q_\nu}{q^2}\right) \ ,
\end{equation}
where the non-perturbative extra piece for the vertex (that, being UV suppressed, will not induce extra
UV divergences) is included under the integral.
Writing
\begin{equation}\label{ghgl6}
\frac{W(q^2)-1}{a^2} \ = \ \frac{1}{b^4+q^4}=\frac{\beta_+}{q^2+ib^2}+\frac{\beta_-}{q^2-ib^2} \ , \qquad\text{with } \beta_\pm=\pm\frac{i}{2b^2}
\end{equation}
one eventually finds after some algebra for \eq{ghgl1} the expression \cite{Cucchieriprogress}
\begin{eqnarray}\label{4D4}
\sigma(k^2) &=& z_0 \ g^2 N_C \, \left\{ \rule[0cm]{0cm}{0.4cm}
\alpha_+ f(k^2,\omega_+^2)+\alpha_- f(k^2,\omega_-^2) \right. \\
&+&  a^2 \left[ \rule[0cm]{0cm}{0.4cm}
\alpha_+ \beta_+ I(k^2,\omega_+^2,ib^2) + \alpha_+ \beta_- I(k^2,\omega_+^2,-ib^2) \right. \nonumber
\\
&+& \left.\left.\alpha_-\beta_+I(k^2,\omega_-^2,ib^2)+\alpha_-\beta_-I(k^2,\omega_-^2,-ib^2)
\rule[0cm]{0cm}{0.4cm} \right]\right\} \ ,
\nonumber
\end{eqnarray}
where\,\footnote{It is assumed that $\text{Re}\left[\omega^2\right] \geq 0$, next to $\nu^2\in i\mathbb{R}$.}
\begin{eqnarray}\label{ghgl8}
I(k^2,w^2,\nu^2) &=& \frac{3}{32\pi^2\epsilon} \ - \ \frac{1}{128\pi^2k^4\omega^2} \\
&\times& \left\{\vphantom{\text{ArcTan}\frac{k^2-\nu ^2+\omega ^2}{\sqrt{-k^4-\left(\nu ^2-\omega ^2\right)^2-2 k^2 \left(\nu ^2+\omega ^2\right)}}}-2k^2\omega^2(5k^2-2\nu^2+\omega^2)-(\omega^6-3\nu^2\omega^4+3k^2\omega^4+3\nu^4\omega^2-3k^4\omega^2)\ln\frac{\omega^2}{\nu^2}\right.\nonumber\\
&&\left.-(k^2+\nu^2)^3\left(\ln\frac{k^2+\nu^2}{\omega^2}+\ln\frac{k^2+\nu^2}{\nu^2}\right)+6k^4\omega^2\ln\frac{\nu^2}{\omu^2}
\right.\nonumber\\
&&\left.+2\sqrt{-k^4-(\nu^2-\omega^2)^2-2k^2(\nu^2+\omega^2)}\left((k^2+\nu^2)^2+2\omega^2(k^2-\nu^2)+\omega^4\right) \right.\nonumber
\\
&& \times \left(\text{ArcTan}\frac{k^2-\nu ^2+\omega ^2}{\sqrt{-k^4-\left(\nu ^2-\omega ^2\right)^2-2 k^2 \left(\nu ^2+\omega ^2\right)}} \right. \nonumber \\
&& \left. \left. + \text{ArcTan}\frac{k^2+\nu ^2-\omega ^2}{\sqrt{-k^4-\left(\nu ^2-\omega ^2\right)^2-2 k^2 \left(\nu ^2+\omega ^2\right)}}\right)\right\}\nonumber
\end{eqnarray}
and we recall from the one-loop computation, with the tree-level vertex, in ref.~\cite{Cucchieri:2012cb},
the function:
\begin{eqnarray}\label{4D8}
    f(k^2,\omega^2)&=& \frac 1 {64 k^4\omega^2 \pi ^2} \ \left\{
    k^4 (-k^2+\omega^2) \ln\left[\frac{\omega^2}{k^2}\right]-\left(k^6-k^4\omega^2+3 k^2\omega^4+\omega^6\right) \ln\left[\frac{k^2+\omega^2}{\omega^2}\right] \right. \nonumber \\
&& \left.  \ + \ k^2\omega^2 (5 k^2+\omega^2+k^2 \ln\frac{k^2}{\mu^2}-4 k^2\ln\frac{k^2+\omega^2}{\mu^2}) \right\} \ ;
\end{eqnarray}
where everything has been computed in the $\msbar$ scheme at the subtraction point $\mu$.
We now have all the ingredients to derive the ghost dressing function in the usual MOM scheme in $d=4$.
To this purpose, we need to properly renormalize the gluon propagator in the MOM scheme\,\footnote{This condition implies a particular requirement for the overall factor, $z_0$, in \eq{ghgl2}.},
\beq\label{momgcond}
\left.D^{\mom}(p^2)\right|_{p^2=\mu^2}=\frac{1}{\mu^2} \ ,
\eeq
and we simultaneously impose
\beq
\left.F^{\mom}(p^2)\right|_{p^2=\mu^2}\ = \ 1 \ ,
\eeq
that can be achieved by the following (one-loop) subtraction again at the renormalization momentum $\mu$:
\begin{equation}\label{4D11}
F^{\mom}(k^2)=\frac{1}{1-\sigma(k^2)+\sigma(\mu^2)} \ ;
\end{equation}
where the ghost self-energy, $\sigma(k^2)$, in r.h.s. is given by \eq{4D4} with $z_0$ determined by the
MOM condition~(\ref{momgcond}) and after replacing the coupling $g$ by the one renormalized in the
so-called Taylor scheme (see for instance ref.~\cite{Boucaud:2008gn}).
We will come back to the latter in the next subsection.

\subsection{Numerical approach to the ghost dressing function}

This section is devoted to deal with the computation of the ghost dressing function by the numerical
resolution of the gap equation, \eq{ghgl0}, through a convergent  iterative procedure. To this purpose,
the bare ghost self-energy can be rewritten as
\begin{equation}
\label{sigmabare}
\sigma(q^2) \ = \  g_0^2 N_c \int \frac{d^4 q}{(2\pi)^4}
\ H_1(q,k) \
\frac{F(q^2)}{(q-k)^2} D((q-k)^2)
\left( \rule[0cm]{0cm}{0.6cm}
1 - \frac{(k\cdot q)^2}{q^2 k^2}
        \right) ,
\end{equation}
after the appropriate kinematical change ($q \to k-q$) for the integration variable
and particularization for $d=4$. Then, after applying the MOM renormalization prescription to
both ghost and gluon propagators, as in \eq{4D11}, the ghost gap equation reads
\beq\label{SD1}
\frac 1 {F^{\mom}(k^2)} \ = \ 1 - \left( \rule[0cm]{0cm}{0.4cm} \sigma(k^2) - \sigma(\mu^2) \right)^{\rm MOM}
\eeq
with
\beq\label{sigmamom}
\left( \sigma(k^2) - \sigma(\mu^2) \right)^{\rm MOM} \ = \
\widetilde{Z}^2_3 Z_3 \frac{g_0^2}{4\pi} N_c \
\int q^3 dq \ K(q,k) \ F^{\mom}(q^2)
\eeq
and the kernel given by
\beq\label{kernel}
K(q,k) \ = \ \frac 1 {\pi^2} \ \int_0^\pi \sin^4{\theta} d\theta \
H_1(q,k) \ \left( \frac{D^{\mom}((q-k)^2)}{(q-k)^2} -
\frac{D^{\mom}((q-p)^2)}{(q-p)^2} \right) \ ;
\eeq
where the subtraction procedure, as detailed in \cite{Boucaud:2008ji}, is applied for two external momenta,
$k$ and $p$, chosen parallel and such that $p^2=\mu^2$, and $Z_3(\widetilde{Z}_3)$ is the gluon (ghost)
MOM renormalization constant. It should be noted that
$H_1$ in \eq{kernel} is a bare but
finite~\cite{Taylor:1971ff} quantity which needs no renormalization
while, in front of the integral, the renormalization constants and the
bare coupling especially appear in the right combination to cancel the
UV singularities and give the MOM T-scheme coupling
(e.g. see \cite{Boucaud:2008gn}),
\beq
\alpha_T(\mu^2) \ = \ \frac{g_0^2}{4\pi} \widetilde{Z}_3^2(\mu^2)
Z_3(\mu^2) \ .
\eeq
This coupling $\alpha_{\rm T}(\mu^2)$ in Yang-Mills QCD is very well known from lattice data
(e.g. see \cite{Boucaud:2008gn}). Thus, we will solve \eq{SD1} by plugging
the gluon propagator, modeled quite precisely with RGZ and lattice QCD as decribed in sec.~\ref{sec:gluon},
into \eq{kernel} to fully determine the kernel for the ghost DSE. Thus,
we can solve the ghost DSE in isolation ({\it i.e.}, without coupling it to the much more
complicated gluon DSE) with the ghost-gluon form factor $H_1$ as the only unknown quantity.

In the previous analytical study, we plugged into the ghost self-energy a simple ansatz for the
ghost-gluon form factor grounded on the OPE analysis of ref.~\cite{Boucaud:2011eh}.
Here, we will also consider the more general formula,
\beq\label{eq:W2}
H_1(q,k) \simeq H_1(q,0) \equiv W(q^2) = c \left( 1 + \frac{a^2 q^2}{q^4 + b^4} \right)
\ + \ (1-c) \ \frac{w^4}{w^4+q^4} \ ;
\eeq
also grounded on the same OPE results, where estimates for $b$ and
$a^2$ defined by \eq{eq:a2A2} can be borrowed from~\cite{Boucaud:2011eh,Boucaud:2008gn}.
As explained in \cite{Boucaud:2011eh}, $c$ is a dimensionless parameter related to the
perturbative contribution to the form factor $H_1$, which is usually assumed to be constant.
The second term in \eq{eq:W2}'s r.h.s. is included to keep the condition
$H_1(0,0)=1$, which appears to emerge, at least approximately, from 4D SU(2) lattice
data~\cite{Maas:2006qw}. Indeed, \eq{eq:W2} is seen to capture the main features
of the current (rough) lattice data for the ghost-gluon vertex. It should be also
noticed that the simpler form factor used in the previous subsection, \eq{ghgl3},
will be recovered from \eq{eq:W2} as $c=1$.

Thus, \eq{ghgl3} and \eq{eq:W2} can be applied in \eq{sigmabare} to solve the gap equation (\ref{SD1})
with the lattice gluon input, and the parameters $c$ and $w$ taken as free parameters
that must be sized to properly account for ghost dressing
lattice data. This will be done in next subsection.

\subsection{Results and comparison}
\label{results}

Let us now discuss the solutions of the GPDSE for the two classes of ghost-gluon vertices (\ref{ghgl3})
and (\ref{eq:W2}) considered here. Recall that both vertices come from the OPE analysis of the corresponding
pure Yang-Mills vertex~\cite{Boucaud:2011eh,Boucaud:2008gn}. As will be shown, the first vertex, being an
approximation to the second vertex by neglecting the perturbative contribution at the subtraction point ($c=1$),
implies that the OPE condensate (included in the parameter $a^2$) to mimic this contribution, resulting in an enhanced value for $a^2$. Our results are summarized in Fig.~\ref{fig:gSDE} and Tab.~\ref{tab:W2}.

In the following, when computing the ghost self-energy,
we take $\alpha_T(\mu=3.6)=0.25$ from ref.~\cite{Boucaud:2008gn}. Furthermore, the different outcomes
of the GPDSE are confronted to the ghost dressing lattice data published in ref.~\cite{Bogolubsky:2009dc},
simulated at $80^4$ and $64^4$ lattices at $\beta=5.7$.

As a self-consistency check, we have found it worthwhile to display also in each case the vertex used, to facilitate a comparison
with, for instance, the SU(3) vertex lattice data of ref.~\cite{Sternbeck:2005re}  or future more precise results.

\subsubsection{Case 1: a trivial vertex}
Our first results are those of the fully iterated GPDSE for the usually made trivial approximation $H_1=1$ (named H10) and in addition, we also consider a ``boosted by hand'' constant behaviour for the form factor (named H1cte); see in Fig.~\ref{fig:gSDE} and Tab.~\ref{tab:W2}. We observe that, as was pointed in refs.~\cite{Boucaud:2008ji,Boucaud:2011ug}, neither approximations are able
to properly account for the behaviour of lattice data. Indeed, the tree level vertex H10 shows deviations
for momenta below 2 GeV, while the H1cte ghost dressing function, although being closer to the lattice data,
also shows clear deviations from the corresponding lattice function.

For comparison, we have also shown in Fig.~\ref{fig:gSDE} the once iterated approximation which, as we have explicitly checked, corresponds one-loop perturbation theory using the RGZ formalism, with the same vertices H10 and H1cte (labeled as H101iter and H1cte1iter). Again, deviations form the lattice ghost
data in the infrared region are striking.

\subsubsection{Case 2: RGZ and NP solutions with vertex model \eqref{ghgl3} }

As a next approximation, we take into account the simple vertex model \eqref{ghgl3} (or equivalently the vertex
model \eqref{eq:W2} with $c=1$) that can describe reasonably well the vertex lattice data.
We will reconsider both the single iteration (named RGZ) and non-perturbative (NP) results of the GPDSE to determine, in each case,
the optimal $a$ and $b$ to the ghost propagator data by treating them as fitting parameters. As a cross-check, for
the single interation results, we applied both the numerical integration and the closed form expression \eqref{4D4}
and found no significant discrepancy (less than 0.5 \% and only for very low momenta, $< 0.03$ GeV).
The results can be found in Tab.~\ref{tab:W2}. Again for the
sake of comparison, we also plotted in Fig.~\ref{fig:gSDE} the results for the single iteration procedure with optimal $a$ and $b$
corresponding with NP results  (labeled as NP1iter).

The vertex \eqref{ghgl3} with the derived value of $a$ and $b$ is shown in Fig.~\ref{fig:vertex}. It is instructive, using the optimal $a$ of both cases, to see what value of $\VA_\s{\rm OPE}$ comes out, something that can be achieved using the OPE motivated formula \eqref{eq:a2A2}. We get $g^2\VA_\s{\rm OPE}\simeq 22.6$ GeV$^2$ for NP and $g^2\VA_\s{\rm OPE}\simeq 23.9$ GeV$^2$
for RGZ which are both unnaturally large when compared to the result of the OPE analysis of Taylor coupling in
ref.~\cite{Boucaud:2008gn}. This is a manifest consequence of the neglection of the perturbative contribution to the
ghost-gluon vertex at the subtraction point (choice $c=1$) which is effectively borrowed by the OPE condensate.
It should also be noticed that, for obtaining a reasonably proper description of lattice data with the single iteration
results, the vertex needs to get quite strongly enhanced at low momenta, mainly due to the small size of the
parameter $b$. \\\

To get an idea what the role is of fully iterating the GPDSE compared to the single iteration of the previous case, we have also shown the once iterated version using the same vertex (and the same parameters) as for the NP case (labeled with NP1iter). Manifestly large discrepancies start to appear for momenta below around 2 GeV and imply that the higher order corrections to the one-loop RGZ ghost propagator are relevant when precise results are the objective. The same could be concluded by comparing H10 and H101iter cases
in the left plot of Fig.~\ref{fig:gSDE} and H1cte and H1cte1iter in the right plot  (solid and dotted blue lines).

\subsubsection{Case 3: fully iterated GPDSE with vertex model \eqref{eq:W2}}

\begin{figure}[htb]
\begin{center}
\begin{tabular}{cc}
\includegraphics[width=7.75cm]{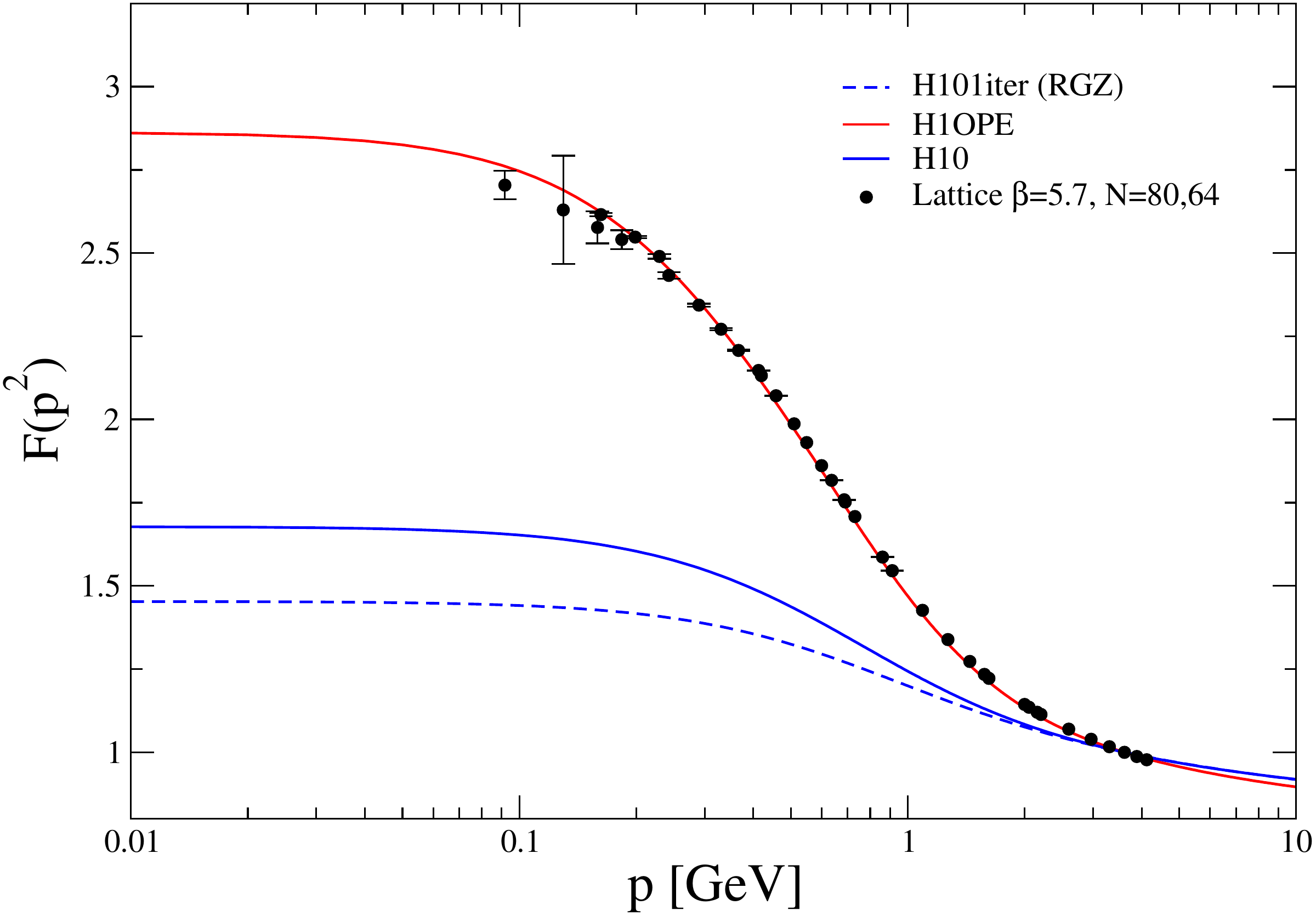} &
\includegraphics[width=7.75cm]{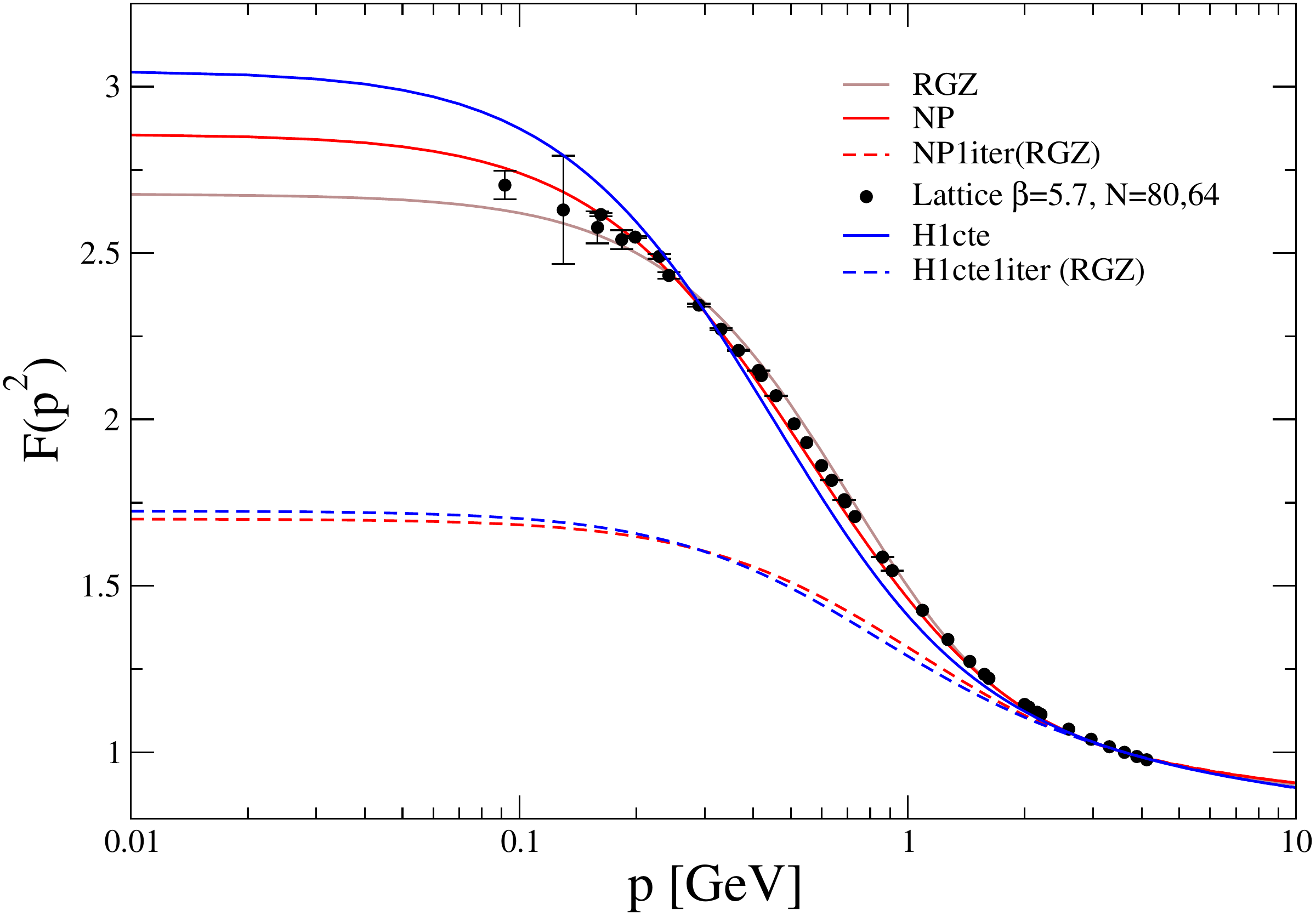}
\end{tabular}
\end{center}
\caption{\small (Left) ghost dressing data obtained by solving the ghost DSE, \eq{SD1}, with a ghost self-energy
given by \eq{sigmamom} and the gluon propagator predicted from \eq{eq:gluonR} and fitted to lattice data
as input for the kernel in \eq{kernel}. The results for the different set of parameters for the form
factor $H_1$ collected (and labeled) in Tab.~\ref{tab:W2} are confronted to the ghost dressing lattice data taken
from ref.~\cite{Bogolubsky:2009dc}. (Right) the same but applying the analytical perturbative approach described
in subsec.~\ref{subsec:anal}. The meaning of the labels for the different curves is explained in the main text.}
\label{fig:gSDE}
\end{figure}

\begin{figure}[htb]
\begin{center}
\includegraphics[width=10cm]{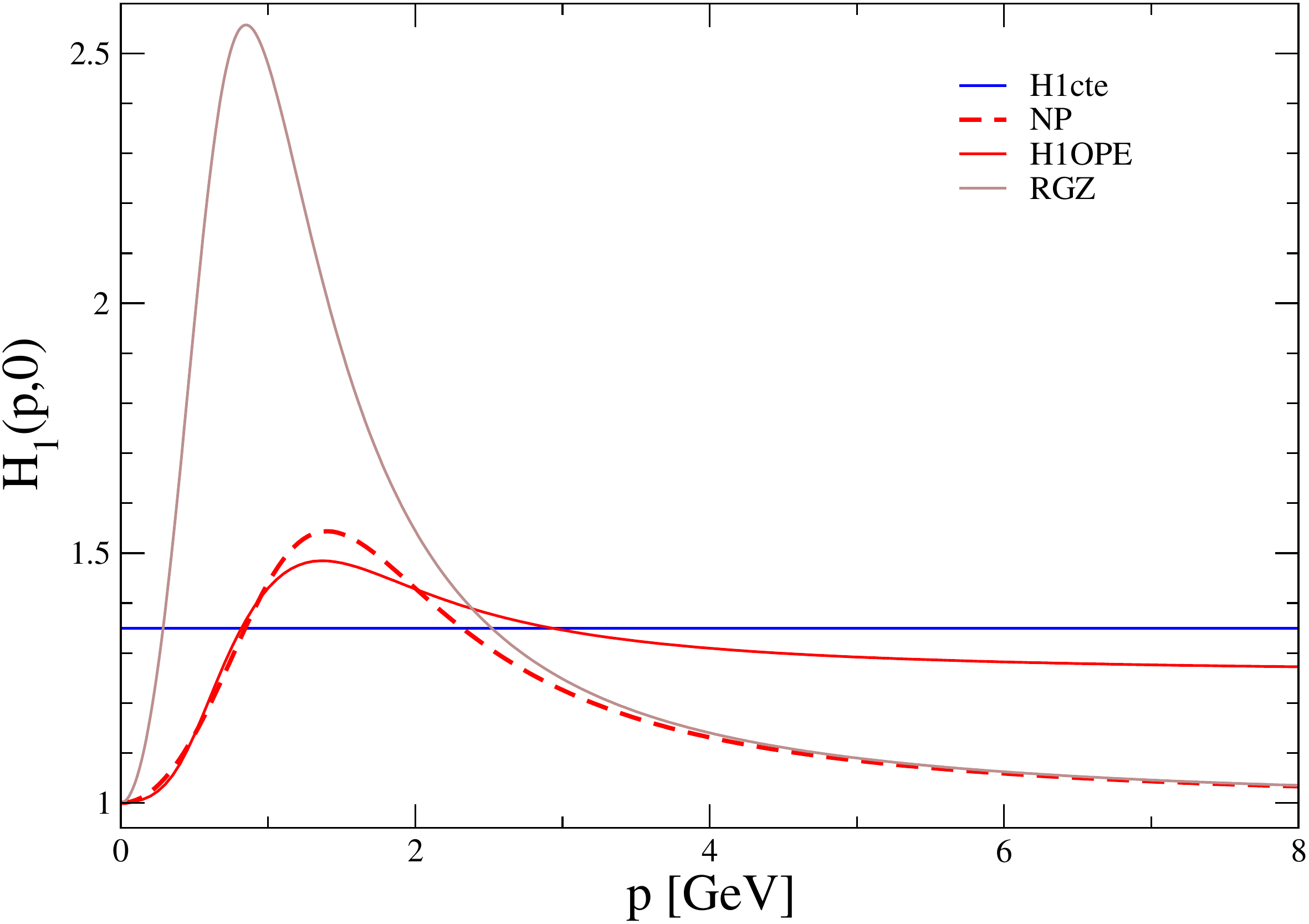}
\end{center}
\caption{\small The different vertex models used to obtain the numerical or the single iteration results for
the ghost dressing function. They correspond to \eq{eq:W2} where the different set of parameters, labeled as
in Fig.~\ref{fig:gSDE}, should be read off from Tab.~\ref{tab:W2}.}
\label{fig:vertex}
\end{figure}

With the full solution of the GPDSE using the somewhat more general vertex model \eq{eq:W2},
our primary goal is to present both a better description of lattice data and relate as much as possible
the vertex parameters to independent analysis in order to give them some ``physical''
meaning. Then, as the full iteration makes it rather difficult to treat all parameters as to be fitted,
we opt to borrow $a$ from the relation \eq{eq:a2A2} and we take $g^2 \langle A^2 \rangle_\s{\rm OPE} \simeq 7$ GeV$^2$,
at tree-level and our renormalization point $3.6$ GeV, from ref.~\cite{Boucaud:2008gn}. We then look for the optimal
values of $c$ and $w$, by the minimization of $\chi^2/d.o.f.$ when fitting the ghost DSE prediction to
the ghost dressing lattice data,  for several values of $b$ around the SU(2) one reported in
ref.~\cite{Boucaud:2011eh} and then we simply took the best $b$. We notice that changing $b$ corresponds
to either shifting the vertex peak towards the large-momentum region and reducing its height or shifting
it towards the low-momentum and increasing its height, thence provided that $a^2$, i.e.~$g^2\VA$, is fixed, the value of
$b$ should not change too much. Fig.~\ref{fig:gSDE} illustrates clearly that this GPDSE solution (labeled H1OPE) captures the lattice data over the full momentum range very well, with the parameters shown in Tab.~\ref{tab:W2}.
Also for the vertex itself, see Fig.~\ref{fig:vertex}, we observe a good resemblance\,\footnote{It should be noticed that the authors of
ref.~\cite{Sternbeck:2005re} plotted $Z_1^{-1}$, which corresponds to the bare but finite form factor $H_1$, rescaled to be 1 at 3 GeV (and dropped any possible large-momentum purely perturbative contribution away, as the one coded by $c$ in \eq{eq:W2}). Although, for a proper comparison with our vertex results, we would have required the non-rescaled bare form factor, but one can anyhow realize that the kinematical structure of their form factor matches pretty well with our result.} with the SU(3) lattice vertex of
ref.~\cite{Sternbeck:2005re} (see Fig.~4 of this paper) or with the SU(2) results of \cite{Maas:2006fk}.

\begin{table}[htb]
   \begin{center}
   \begin{tabular}{l@{\hspace{0.5cm}}l@{\hspace{0.3cm}}l@{\hspace{0.3cm}}l@{\hspace{0.3cm}}l@{\hspace{0.3cm}}l}
   \hline
   label & $c$    &  $a$  & $b$  & $w $ & $\chi^2/d.o.f.$ \\
         &        & (GeV) &(GeV) &(GeV) &                \\	
   \hline
     H1OPE & 1.26 &  0.80  & 1.3  & 0.65 & 0.83   \\
     H1cte & 1.35 &  0	  &  -	 &  -   & 13.0   \\
     \hline
   	 NP   & 1 & 1.46 & 1.4 & - & 1.74 \\
     RGZ & 1 & 1.50 & 0.85 & - & 1.98 \\ 	
   	\hline
     H10   & 1    &  0    &  -   &  -   &  -     \\
   \hline
   \end{tabular}
   \end{center}
      \caption{\small Parameters for the form factor $H_1$, given by \eq{eq:W2},
      used to solve the ghost DSE.  The set of
      parameters are refereed in Fig.~\ref{fig:gSDE} as labeled in the first column. The first three lines correspond
      to fits over the whole momenta window, while the fourth stands only for momenta above 0.3 GeV.}
    \label{tab:W2}
\end{table}

\section{Conclusion}
As the main goal of this paper, we have shown the quantitative compatibility between large-volume lattice ghost propagator and its DSE description. To this purpose, we discussed
the relevant role played by a non-trivial ghost-gluon form factor (in particular, the non-longitudinal form factor usually
called $H_1$) that has been recently modeled founded in the OPE vertex description and in good
agreement with current (although not yet very precise) vertex lattice results. Indeed, we have shown that,
only after the appropriate inclusion of such a non-trivial form factor, the integration of the gluon propagator
in the ghost DSE will result in a ghost propagator numerically consistent
with the large-volume lattice data. A key point in our analysis is the fact that, once the gluon propagator is
precisely borrowed from the lattice and the renormalized coupling unambiguously related to
the one in Taylor scheme, the only unknown DSE ingredient is the above form factor.

As an easy-to-handle model for the gluon propagator is of
a great help to solve the ghost DSE, after being properly fitted to lattice data, we invoked the tree-level gluon
propagator within the RGZ approach. This RGZ gluon propagator fulfills a two-sided goal: it provides, after applying the
usual MOM renormalization prescription, with an appropriate model describing lattice data to be plugged into the DSE
and solve it numerically; and, being so simple as a model, it allows for an approximate analytical result based on the
one-loop RGZ approach.  Then, the results are compared and we concluded
that the perturbative RGZ approach (= 1st iteration of GPDSE) allows for a good description of the non-perturbative lattice results only after one considerably enhances the ghost-gluon vertex at low momenta, a fact not really supported by the lattice data
of \cite{Sternbeck:2005re}.

\bigskip

{\bf Acknowledgements:}
 J.~R.-Q.~acknowledges the Spanish MICINN for the support by the research project FPA2011-23781 and ``Junta de Andalucia'' by P07FQM02962. D.~D.~is supported by the Research-Foundation Flanders (FWO Vlaanderen). O.~O.~acknowledges financial support from the F.C.T.
research project PTDC/FIS/ 100968/2008,  developed under the initiative QREN financed by the UE/FEDER through the Programme COMPETE - Programa Operacional Factores de Competitividade. 
We wish to thank A.~Cucchieri, O. P\`ene, Ph. Boucaud and J.P.~Leroy for inspiring discussions. We are furthermore grateful to the Berlin, Moscow and Adelaide lattice groups for providing us with their data.


\providecommand{\href}[2]{#2}\begingroup\raggedright\endgroup

\end{document}